\documentstyle[11pt,paspconf,epsf]{article}
\begin{document}

\title{Peculiar Motions and the Galaxy Density Field}

\author{Christian Marinoni, Giuliano Giuricin, Barbara Costantini}
\affil{
Dept. of Astronomy, Trieste University and SISSA, Trieste, Italy}
\author{Pierluigi Monaco}
\affil{
Institute of Astronomy, Cambridge, UK}

\begin{abstract}
We use an all--sky, complete sample of nearby galaxies, extracted 
from the LEDA data base, to map the optical galaxy density field 
in the nearby universe. In order to determine this field, we correct 
the redshift--dependent distances by testing some peculiar velocity 
field models and we correct the galaxy number density for the 
incompletion of the galaxy sample at large distances through the 
derivation of the galaxy luminosity function.  Local galaxy density 
parameters calculated for different smoothing scales are meant to 
be used in forthcoming statistical studies of environmental 
effects on galaxy properties. 
\end{abstract}

\keywords{galaxies: distances and redshifts -- cosmology: large-scale 
structure of the universe
}

\section{Introduction}

Optical galaxy samples are more suitable than IRAS--selected galaxy
samples for mapping the galaxy density field on quite small scales ($ < 1$
Mpc). In our work, we attempt to recover this field using a 3D
recontruction procedure (based on peculiar velocity field models) which
minimizes relevant systematic effects. 

To this aim, we take an all--sky, magnitude--limited, optical sample 
of nearby galaxies (6392 galaxies with recession velocities 
$cz<5500$ km/s) extracted from the LEDA data base (Garcia, 1993). 
This sample, which is complete up to the corrected blue total 
magnitude $B_T=14$ mag, comprises 3403 field galaxies and 485 
systems with at least three members (for a total of 2989 galaxies).
 
In order to correct raw redshift--distances, we use two basic models of
the peculiar velocity field: i) an optical cluster 3D--dipole
reconstruction scheme devised by Branchini \& Plionis (1996) that we
modify with the inclusion of a local model of the Virgocentric infall; ii)
a multi--attractor model, in which we adopt a King density profile for
each attractor (i.e., the Virgo cluster, the Great Attractor, the
Perseus--Pisces and Shapley superclusters) and we rely on the weakly
non--linear peculiar series expansion by Reg\"os \& Geller (1989) for
peculiar velocity. We fit the multi--attractor model to the Mark II and
Mark III peculiar velocity catalogues (Willick et al., 1997). 

Throughout we adopt the Hubble constant $H_0=100\; km\; s^{-1} Mpc^{-1}$. 
 
\section{Results} 

Interestingly, as a result of the non--degenerate manner in which the
cosmological density parameter $\Omega_0$ enters into the equations of
motions, we find the best-fitting value of $\Omega_0=0.5\pm0.2$ for our
multi--attractor model fitted on Mark III data (whilst Mark II data would
yield $\Omega_{0}\sim 1$). This is consistent with a picture in which a
large part of the local peculiar flows ($ < 8000 $ km/s) is generated on
a scale which is larger than the analyzed volume (from distant gravitational
sources such as the Shapley concentration). Moreover, for the Great
Attractor and Perseus--Pisces, the mass overdensity peaks, evaluated from
our Mark III multi--attractor model at a radial distance of 1200 km/s, are
not much different from the reconstructed overdensity peaks of IRAS
galaxies (Sigad et al., 1997) and optical galaxies (Hudson et al., 1995),
calculated for a Gaussian smoothing of 1200 km/s. This points to a bias
factor $b$ of order unity for optical and IRAS galaxies over a $\sim10$ 
Mpc scale (see Marinoni et al. 1998 for details). 

Inverting the redshift--distance relations predicted by the 
above--mentioned velocity field models, we derive the distances 
of the field galaxies and groups of our sample. We overcome the 
ambiguity inherent to the triple--valued zones of the redshift--distance 
relations by using blue Tully--Fisher relations calibrated on galaxy 
samples having distances predicted by the velocity models.

The use of different velocity field models allows us to check to what 
extent differences in the description of the peculiar flows influences 
the estimate of galaxy distances in the nearby universe. We note 
that these differences turn out to be more prominent at the largest 
and smallest distances rather than for intermediate distances (i.e., 
for $2000 < r < 4000$ km/s, where $r$ is the distance expressed in 
km/s).

We correct the incompletion of the sample at large distances through the
derivation of the Schechter--type blue luminosity function of galaxies
$\Phi (L)$.  The incompletion factor $F(r)$ which expresses the number
of galaxies that should have been catalogued for each objet present in
the sample at a given distance r, is related to the selection function
$\phi(r)$ by the expression $F(r)=1/\phi(r)$. The selection function is
given by $\phi (r)=1$ if $r<r_{s}$, where $r_{s}=5$ Mpc, and is given 
by  
\begin{equation} 
\phi(r)=\frac{\int_{L_{min}(r)}^{\infty}\;\;\Phi(L)dL}
{\int_{L_{s}}^{\infty}
\;\Phi(L)dL}
\end{equation} 

\noindent 
if $r>r_{s}$, where $L_{min}$ is the minimum luminosity
necessary for a galaxy at distance $r$ to be included in the sample. We set
the lower limit of the integral in the denominator to
$L_{s}=L_{min(r_s)}$. 
 
By applying a modified Turner (1979) method, which is insensitive to the presence
of fluctuations in the sample, we derive the parameters $M^{*}$ (the
characteristic magnitude), $\alpha$ (the slope of the faint tail), and
$\Phi^{*}$ (the normalization factor) of this function on the basis of the
distances predicted by velocity field models. Specifically, comparing
the absolute magnitude distribution with that expected on the basis of
the following estimator
\begin{equation}
{\cal N} (M_i) \Delta M=\Phi(M_i) \Delta M \sum_{j=0}^{j_{lim}}\bigg[
\frac{N({\bf r}_j)\Delta r}{\int_{- \infty}^{M_{lim,i}}\;\;\Phi(M^{'})dM^{'}}
\bigg]
\end{equation}

\noindent we obtain $\alpha=-1.06\pm 0.04$, $M^*=-20.12\pm0.06$ and
$\Phi^{*}=1.01\cdot 10^{-2}\; Mpc^{-3}$ for the distances predicted by the
Mark III multi--attractor model and $\alpha=-1.18\pm 0.04$,
$M^*=-20.19\pm0.06$ and $\Phi^{*}=9.91\cdot 10^{-3}\; Mpc^{-3}$ for the
redshift-distances evaluated in the CMB frame. Similar values result from
sets of distances relative to other velocity field models such as the Mark
II multi--attractor model and the cluster dipole model. Thus, within the
current views on the kinematics of local peculiar flows, corrections of
galaxy distances for peculiar motions appear to have a small impact on the
galaxy luminosity function, which in our case is a quantity calculated
over a very large solid angle. 

We devise a method of 3D reconstruction of galaxy groups, which prevents
members to be spuriously placed in high--density regions and we
calculate the local galaxy density $\rho_{\sigma}$ (in galaxies per
$Mpc^{3}$) in terms of the number density of galaxies which are found
around every galaxy. This is done by smoothing every galaxy with a
Gaussian filter having a fixed smoothing scale parameter $\sigma$ (in Mpc)
(see also Giuricin et al., 1993; Monaco et al., 1994): 

\begin{equation}
\rho({\bf r}_{i}) = \frac{1}{(2\pi\sigma^{2})^{3/2}}\cdot 
\sum_{j \neq i}^{} \exp \left( 
\frac{|{\bf r}_{j} - {\bf r_{i}}|^{2}}{2 \sigma^{2}} \right) \, F(r_{j}) 
\end{equation} 

\noindent
This quantity gives the number of galaxies (brighter than the absolute 
magnitude $M_B$=--17.4 mag) per $Mpc^{3}$, which are located within a 
distance equal to $\sim\sigma$ from the specified galaxy of the sample; 
F is the correction function for the incompleteness of the catalogue 
at large distances. The sum is carried out over all galaxies 
except the one whose density we are calculating.

Choosing different values of the smoothing scale parameter $\sigma$ allows
us to explore the 3D galaxy density field on different high--resolution
scales. In particular, for sets of galaxy distances relative to 
different velocity models, we have calculated the local galaxy density 
$\rho_{\sigma}$, choosing $\sigma$=0.25, 0.5, 1, and 2 Mpc. 

Fig. 1 shows the comparison between the $\rho_{\sigma}$-values based on
redshift--distances (in the LG frame) and on distances predicted by Mark
III, Mark II, and cluster dipole (cd) models, for $\sigma$=0.25 and 2 Mpc.
We also show the comparison between the Mark III $\rho_{\sigma}$-values
for $\sigma$=2 Mpc and the galaxy densities which are deduced from the
IRAS 1.2 Jy redshift survey through smoothing with a Gaussian whose
dispersion is equal to the local mean galaxy separation (Fisher et al.,
1995). On large scales ($\sigma$=2 Mpc), the agreement between the
different sets of $\rho_\sigma$-values is satisfactory at low or
intermediate galaxy densities; at large galaxy densities, Mark III and
Mark II multi--attractor models tend to give greater values than those of
the IRAS galaxy sample and lower values than those of the other models. On
the other hand, in general there is a poor agreement between the various
sets of $\rho_\sigma$-values on small scales ($\sigma$=0.25 Mpc). 

In conclusion, corrections of galaxy distances for peculiar motions 
appear to have a large impact on the evaluation of the local galaxy 
density on small scales ($<1$ Mpc). This is an important parameter 
to be used in statistical studies of environmental effects on 
the properties of nearby galaxies.\\

\begin{figure}[ht]
\plotone{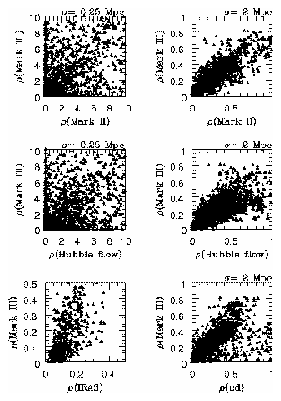}
\caption{Comparison between the local
galaxy densities (in $Mpc^{-3}$) relative to the 3D optical galaxy
distribution based on different velocity field models and to the IRAS 1.2
Jy galaxy sample.} \label{fig-1} \end{figure}

\end{document}